\documentstyle[twoside,fleqn,espcrc2]{article}

\newcommand{\AmS}{{\protect\the\textfont2
  A\kern-.1667em\lower.5ex\hbox{M}\kern-.125emS}}

\title{Isgur-Wise Function on the Lattice}

\author{Claude W. Bernard,\address{Physics Department,
        Washington University,
        St. Louis, MO 63130, USA}
        Yue Shen\address{Physics Department,
        Boston University,
        Boston, MA 02215, USA}
        \thanks{Speaker at the conference}
        and
        Amarjit Soni\address{Physics Department,
        Brookhaven National
        Laboratory, Upton, NY 11973, USA}}

\begin{document}

\begin{abstract}
We review our method and numerical results for calculation of
the Isgur-Wise function on the lattice. We present a discussion of
the systematic errors. Using recent experimental results,
we find $V_{cb} = 0.044\pm 0.005\pm .007$.
\end{abstract}

\maketitle

\section{Method}
A very active subfield in high energy physics recently is the study
of hadrons with heavy-light quark content \cite{Neubert}.
A major effort has been spent
in calculating the Isgur-Wise function, which, once
it is determined, can be
widely used in calculations of heavy meson decay ($b \to c$) processes.
After an initial exploration \cite{BSS1}, calculations of the Isgur-Wise
function on the lattice \cite{BSS2,UKQCD,Kenway} have quickly obtained
interesting results which can be directly compared with experimental data
and can be used to determine one of the elements of
the  CKM matrix, $V_{cb}$,
in the Standard Model.

For calculating the Isgur-Wise function, $\xi(v\cdot v^\prime)$,
on the lattice, we have proposed
\cite{BSS1} to use the flavor symmetry of the heavy quark effective theory
(HQEFT) \cite{Bjorken} and measure the $D \to D$ elastic scattering
matrix element
\begin{equation}
<D_{v^\prime} | {\bar c}\gamma_\nu c|D_{v}> = m_D C_{cc}(\mu)
\xi (v\cdot v^\prime; \mu) (v + v^\prime)_\nu ~,
\label{eq:matrix}
\end{equation}
where $m_D$ is D meson mass, $v$ and $v^\prime$ are
four-velocities of the initial and final D mesons.
The constant $C_{cc}(\mu)$
represents the QCD renormalization effect from the heavy
quark scale to a light
scale $\mu$. The calculation was performed in the quenched approximation using
Wilson fermions. Both light and heavy quarks are treated as propagating.
For details of the
numerical simulation, refer to Refs. \cite{BSS1,BSS2}.

\null From the lattice point of view, calculating the elastic scattering matrix
element has significant advantages. In comparison to the $B \to D$ process,
the elastic process
on the lattice has much less noise and therefore has smaller statistical
errors. Furthermore, because of the exactly known value
\begin{equation}
<D_{v} | {\bar c}\gamma_\nu c|D_{v}> = 2m_D,
\label{eq:norm}
\end{equation}
at the ``zero recoil'' point $v^\prime = v$, the lattice artifacts that are
independent of momentum can be removed without ambiguity using
Eq.~(\ref{eq:norm}) as normalization condition for lattice
data \cite{BSS1,BSS2}. A similar strategy for $B \to D$ decay would have
introduced an extra (unknown)
$O(1/m_Q^2)$ correction.
Therefore, not surprisingly, the most accurate
data obtained so far on the lattice are from $D \to D$ elastic
scattering \cite{BSS2,UKQCD}.
However,
inelastic processes,
such as $B \to D$ and $B \to D^*$, can be valuable consistency
checks \cite{UKQCD,Kenway}.

\section{Systematic Errors}

For analysis of the systematic errors, let us consider the slope, $\rho^2$,
of the Isgur-Wise function at $y \equiv v\cdot v^\prime = 0$. A fit of the
lattice data \cite{BSS2} to the
relativistic harmonic oscillator model \cite{NeuRie}
\begin{equation}
\xi(y) = {2\over y+1} \exp\left[-(2\rho^2_{NR}-1){y-1\over y+1}\right]~,
\label{eq:NR}
\end{equation}
gives
$\rho_{NR}^2 = 1.41(19)$. For a model independent determination
of $\rho^2$,
one may choose to fit $\xi(y)$ near $y=1$
\begin{equation}
\xi (y) = 1 - \rho^2 (y-1),
\end{equation}
and obtain \cite{BSS2}
$\rho^2 = 1.24(26)$. All the fits have taken account of the correlations
between data points using covariance matrices. There are several potential
sources of systematic corrections: quenching,
scaling violation, light quark
mass $m_q$ dependence, finite volume effect,
heavy quark mass $m_Q$ dependence.

{\it Quenching.} The error due to quenching is the most difficult to quantify.
Although the effect is expected to be small if a scale such as $f_\pi$ is set
to the physical value
(we use Ref.~\cite{BLS} to set the scale with $f_\pi$),
a systematic
numerical study is still lacking. We will not give an assessment on the
quenching effect here.

{\it Scaling violation.} Since by using the normalization condition
Eq.~(\ref{eq:norm}) all the momentum independent lattice artifacts are
removed and the remaining scaling violations
are proportional to $y-1$.
Therefore, we expect
the residual scaling violations to be small. A fit to data at $\beta=6.3$
and $\beta=6.0$ found a difference of $13\%$ for $\rho_{NR}^2$.

A direct check on the Euclidean invariance on the lattice is to measure
the ratio of the form factors $f_-/f_+$. This ratio was found small
and consistent with zero within large errors \cite{BSS1,BSS2}.

{\it Light quark mass ($m_q$) dependence.} Our lattice data for $\xi(y)$
are presented with $m_q$ set to the strange quark mass, $m_q=m_s$.
These data are directly relevant to processes such as $B_s \to D_s$,
$B_s \to D_s^*$. For $B \to D$, they have to be extrapolated to
the ``chiral limit'' $m_q = m_{u,d} $.
An inspection
shows that the linear size of the physical volume is in the range of
$(100MeV)^{-1}$ \cite{BSS2}, therefore, at
$m_q < m_s$ the finite size effect becomes important and
contaminates
the $m_q$ dependence.

To estimate $m_q$ dependence
we therefore
use data obtained on the largest physical volume ($24^3\times39$ lattice at
$\beta=6.0$).
We first estimate the shift in $\rho_{NR}^2$ from $m_q$ to $m_{q^\prime}$
with both $m_q, m_{q^\prime}$ in the range of $m_s$. Then this shift
in $\rho_{NR}^2$ is extrapolated to the chiral limit. Using this procedure,
we find $\rho_{NR}^2$ {\it decreases} by $12\%$ from $m_q = m_s$ to
$m_q = m_{u,d}$ \cite{BSS2}. It is
interesting to note that the sign of this
shift is opposite to the chiral perturbation prediction \cite{Jenkins}
and in agreement to the bag model calculation \cite{Sadzi}.
It is important to confirm this trend in the future
with improved statistics.

{\it Finite volume effect.} To estimate the finite
volume effect, we compare
our data on $16^3\times39$ and $24^3\times39$
lattices at $\beta=6.0, \kappa_q=.154$
($m_q=m_s$). There is a shift of $15\%$ in $\rho_{NR}^2$. We expect
that the finite size effect would be smaller at a heavier $m_q$. Indeed,
the shift in $\rho_{NR}^2$ is reduced to $9\%$ at $\kappa_q=.152$.

{\it Heavy quark mass ($m_Q$) dependence.} Recent lattice calculations
indicated that the heavy quark symmetry
begins to set in in the neighborhood of the charm
mass. The leading $1/m_Q$ dependence agrees with the
expectations of HQEFT. We refer to Ref. \cite{BSS2} for discussions of
specific examples. Therefore, simulation results obtained at the
charm mass
range can be used and extrapolated to the heavy quark limit. For the
Isgur-Wise function the leading order $1/m_Q$ correction should be
$\sim (y-1)\Lambda_{QCD}/m_Q$ \cite{Luke}. It should be relatively
small for current lattice calculations $y-1 < 0.2$. Indeed, comparing
$\rho_{NR}^2$ at $m_Q \sim 1.6$GeV and $2.3$GeV, we find $15\%$ shift.

{\it Summary.} Adding up the above items in quadrature,
the total systematic correction becomes $29\%$. We have
\begin{equation}
\rho_{NR}^2 = 1.41 \pm .19 \pm .41,
\label{eq:NRfit}
\end{equation}
where the first error is statistical and the second is systematic error.
For linear fit, we get
\begin{equation}
\rho^2 = 1.24 \pm .26 \pm .36,
\label{eq:Lfit}
\end{equation}

We should point out that this $29\%$ systematic error is probably an
overestimate. Our fit in Eqs.~(\ref{eq:NRfit}) and (\ref{eq:Lfit})
have been performed with data at all $\beta$, lattice size, heavy quark
mass values. Therefore, the combined systematic error is
unlikely to be much
larger than the statistical error ($.19$). Indeed,
though we use it as an indication of the systematic errors,
the shift in $\rho^2$ due to
each item discussed above is not statistically significant.
To get a better
analysis of the systematic errors, one needs more data points and
better statistics.
At this point,
our discussion of the systematic errors should be taken
primarily as a discussion on the methodology; the estimates
obtained are only qualitative.

A comparison with continuum model calculations is given in Ref. \cite{BSS2}.
Clearly, lattice result has reached similar, if not better,  numerical
accuracy as the continuum models for $\rho^2$.

Our result for $\rho^2_{NR}$ is also consistent with a recent lattice
calculation by UKQCD Collaboration \cite{UKQCD,Kenway}.

\section{Extracting $V_{cb}$}

Although $\rho^2$ is useful for comparison with continuum model calculations,
it is less useful for getting $V_{cb}$ from the experimental data. Around
$y =1$, experimental data have the lowest statistical precision \cite{ARGUS}.
On the lattice, a model independent determination of $\rho^2$ also tends to
have larger uncertainty because only a few data point close to $y=1$ can be
used.

\begin{figure}[htb]
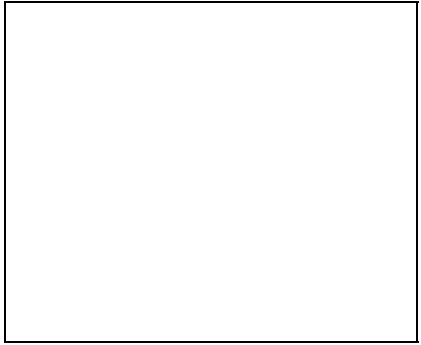

\vspace{9pt}
\framebox[55mm]{\rule[-21mm]{0mm}{43mm}}
\caption{Comparison of
present lattice data (crosses) with experimental data (open
circles) [7].
The solid line is a fit to the lattice data according to
Eq.~(3).}
\end{figure}

In ARGUS (and CLEO) experiments, what has been measured is $|V_{cb}|\xi(y)$
with the most accurate data obtained in the
range $1.1 < y < 1.5$. In the lattice
calculation, we have
obtained $\xi(y)$ in the range $1 < y < 1.2 $. Therefore,
at least in the range $ 1.1 < y < 1.2$ we can directly fit the experimental
data with the lattice data with only one unknown parameter $V_{cb}$.
One such fit is shown in Fig. 1. We obtain \cite{KEK}
\begin{equation}
|V_{cb}|\sqrt{\tau_B\over 1.53ps} = 0.044 \pm .005 \pm .007,
\end{equation}
where the first error is due to the statistical and systematic errors in the
lattice calculation and the second error is from the experimental
uncertainties. The errors on the lattice data essentially reflect the
spread of $\xi$ over different $\beta$, lattice size, and heavy quark mass
values.

\end{document}